\documentclass{desyproc}
\begin{document}
\title{Statistical criteria for possible indications of new physics in tritium $\beta$-decay spectrum}

\author{{\slshape Aleksei Lokhov$^1$, Fyodor Tkachov$^2$}\\[1ex]
$^1$Dept. of Experimental Physics, Institute for Nuclear Research RAS, 142190 Troitsk, Russia\\
$^2$Dept. of Theoretical Physics, Institute for Nuclear Research RAS, 117312 Moscow, Russia}

\contribID{xy}

\confID{8648}  
\desyproc{DESY-PROC-2014-04}
\acronym{PANIC14} 
\doi  

\maketitle

\begin{abstract}
The method of quasi-optimal weights is applied to constructing (quasi-)optimal
criteria for various anomalous contributions in experimental spectra.
Anomalies in the spectra could indicate physics beyond the Standard Model
(additional interactions and neutrino flavours, Lorenz violation etc.). In
particular the cumulative tritium $\beta$-decay spectrum (for instance, in
Troitsk-$\nu$-mass, Mainz Neutrino Mass and KATRIN experiments) is analysed
using the derived special criteria. Using the power functions we show that the
derived quasi-optimal criteria are efficient statistical instruments for
detecting the anomalous contributions in the spectra. 
\end{abstract}

\section{Introduction}

Studying anomalies in experimental spectra extends our understanding 
of experimental setups. Besides anomalous contributions could also indicate new
physics beyond the Standard Model. For instance, in tritium $\beta$-decay
spectra possible additional interactions can lead to a step-like anomaly near
the end-point while the excistance of the forth neutrino (with the mass of a few
keV) induces a kink structure in the region of several keV from the end-point.
Here we consider these two possible anomalous contributions.

The search for an anomaly should be based on a statistically reliable
inference about presence or absence of the anomaly. Such inference is provided
by special statistical criteria. One can construct the criteria according to
each particular situation and accounting for some additional information about
the theoretical model or experimental setup. The various approaches here are as
follows:

\begin{enumerate}
 \item Direct fit with additional parameters (the mass of neutrino and the
mixing parameter, the amplitude and the position of the step)
\cite{LobNPA2003}, \cite{AsePRD2011}, \cite{MerArx2014}.
 \item Searching for the kink with various filters \cite{MerArx2014}.
 \item Wavelet analysis \cite{MerArx2014:wavelet}.
 \item Seaching for special functional dependencies.
 \item Constructing special statistical criteria for the heavy neutrino or the
step accounting for the uncertainties of other parameters (based on the method
of quasi-optimal weights \cite{TkaArx06}). In the paper we present two
examples of construction of the special statistical tests.
\end{enumerate}

\section{Step-like anomaly in Troitsk-$\nu$-mass spectrum}
Any deviations at the very end of the spectrum have crucial influence
on the estimations of the neutrino mass squared. The spectrum of the
Troitsk-$\nu$-mass experiment is cumulative. The
numbers of electrons are measured for a set of energies (these
numbers have Poisson distributions). After that the measured data
points are fitted with the theoretical curve. There are four fitted
parameters in the Troitsk-$\nu$-mass spectrum. One of them is
the neutrino mass squared.

The first data analysis \cite{LobNPA2003}, performed with the standard Minuit
routines, yielded rather controversial result: the estimate of the
neutrino mass squared lies far beyond physically relevant range, it appears to
be large and negative.
This was interpreted as due to an excess of electrons near the end-point energy
of the tritium $\beta$-decay spectrum; in cumulative spectra,
such an excess takes the form of a step.
Such a step is described by two parameters, the height and the position.
Including these into the fit, a satisfactory value for the neutrino mass
squared was obtained.

The recently finished new analysis \cite{AsePRD2011} (exploiting the method of
the quasi-optimal weights \cite{TkaArx06} and improved theoretical model of the
experimental setup as well), yielded physically relevant values (within
errors) of the neutrino mass squared while the step-structure has not
been accounted for.
The goodness-of-fit test included into the fitting procedure is not
tuned to feel the anomalous contributions of this step-like form. It will
be nice to have convenient, robust statistical criteria, particularly
targeted to the described anomaly. We also should take into account
that the position of the step is unknown and even may vary in time.
We have constructed three special criteria and with them one can
perform the standard procedure of the test of hypotheses
\cite{LokTkaTruNIMA2012}, \cite{LokTkaTruNPA2013}. The
null-hypothesis is that the height of the step is zero, the alternative -
the height is positive. We use the fit from the new analysis \cite{AsePRD2011}.

The first criterion is constructed via routines of the method of quasi-
optimal moments. And it is by construction the Locally Most
Powerful (LMP) one. Locally here means near the null-hypothesis.
The distributions for the experimental counts are $f_i(N)=\frac{\mu_{i}^{\prime
N}e^{-\mu^{\prime}_i}}{N!}$, where $ \mu_{i}^{'}\rightarrow\left\{
  \begin{array}{l l}
  \mu_{i}+\Delta_{-},	 i>m \\
  \mu_{i}+\Delta_{+},	 i\leq m
\end{array}
 \right.$, $i$ stands for the number of an experimental point.
Here $m$ is defined by the inequality $E_m\leq E_{st} \leq E_{m+1}.$
Constructing the weights  
$$
 \omega_{i}^{+}(N)=\frac{\partial \ln f_i}{\partial \Delta_{+}}=
 \left\{
 \begin{array}{l c}
  0, & i>m \\
  \frac{N}{(\mu_i+\Delta_{+})}-1, & i \leq m
 \end{array}
 \right.,
 \omega_{i}^{-}(N)=\frac{\partial \ln f_i}{\partial \Delta_{-}}=
 \left\{
 \begin{array}{l c}
  0, & i \leq m \\
  \frac{N}{(\mu_i+\Delta_{-})}-1, & i > m
 \end{array}
 \right.
$$
and solving the corresponding equations
$h^{exp}= \sum\limits_i \omega_i=0$ one obtains the statistics of the LMP
criterion $\overline{\Delta}=\Delta_{+}-\Delta_{-}$ -- the estimate for the
height of the step. $\overline{\Delta}$ can be also presented as a weighted sum
of experimental counts $\overline{\Delta}=\sum_{i=1}^{M}w_i \cdot N_i$

\begin{figure}[h]
\begin{center}
\begin{minipage}[h]{0.3\linewidth}
\includegraphics[width=1\linewidth]{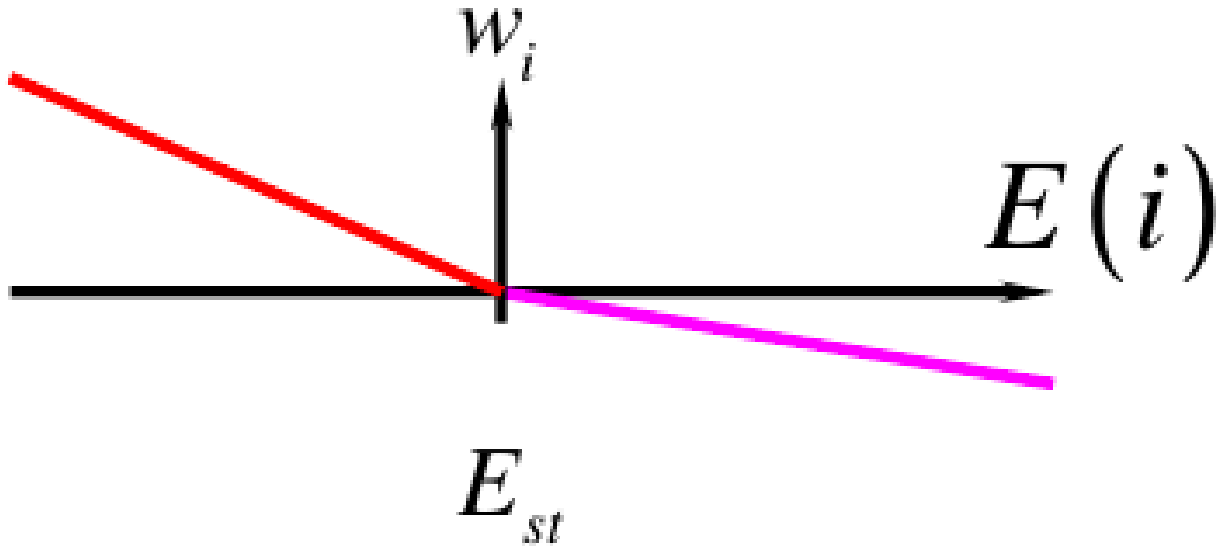}
\caption{The quasioptimal weights for the step-like anomaly searches}
\label{fig:q-opt}
\end{minipage}
\hfill
\begin{minipage}[h]{0.3\linewidth}
\includegraphics[width=1\linewidth]{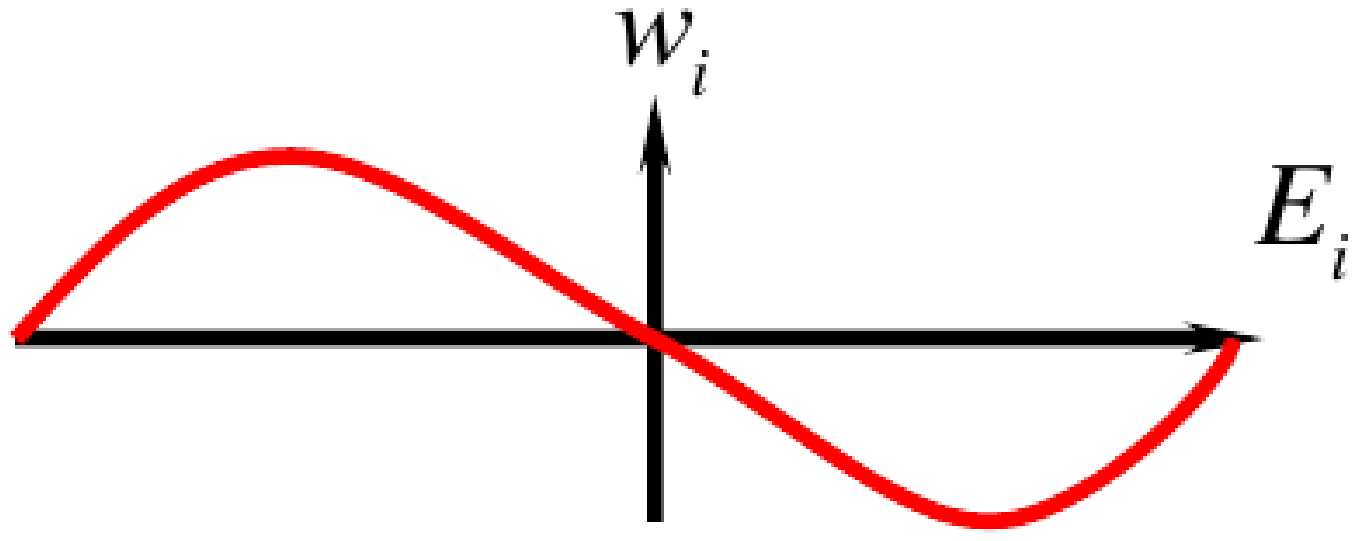}
\caption{The quasioptimal weights for the heavy neutrino searches}
\label{fig:q-opt2}
\end{minipage}
\hfill
\begin{minipage}[h]{0.3\linewidth}
\includegraphics[width=1\linewidth]{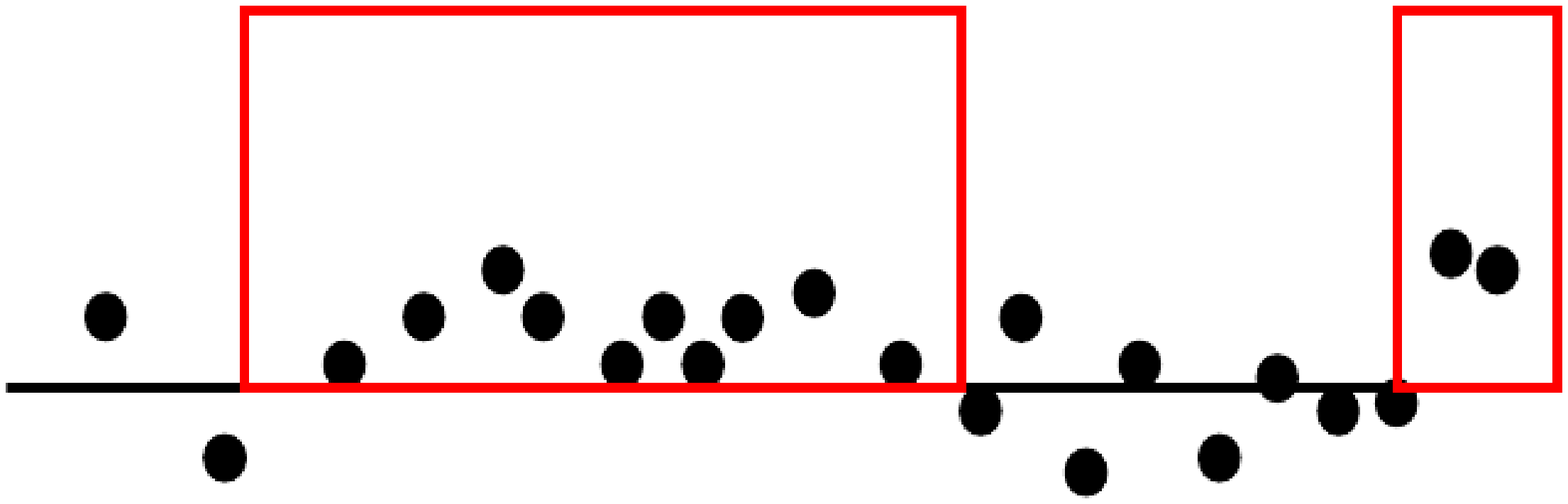}
\caption{The idea of the pairwise neighbours correlations criterion}
\label{fig:pairwise}
\end{minipage}
\end{center}
\end{figure}

Recalling the uncertainty of the step position it is useful to decrease
the sensitivity of our criteria to the position of the step, even loosing
some sensitivity to the step itself. For this we slightly change the
weights in the sum of the LMP test (see Fig. \ref{fig:q-opt}), to suppress the
values near the position of the step, saving the properties of the LMP test in
the rest areas. The corresponding statistics is $ S_{q-opt}= \sum\limits_{i}^{k}
w_i \cdot \xi_i $, where $\xi_i = \frac{N_i-\mu_i}{\sqrt{\mu_i}}$ and
$w_i=\left\{
 \begin{array}{l c}
  \frac{(m-i)}{m}, & i \leq m, \\
  \frac{(m-i)}{M-m}, & i > m.
 \end{array}
\right.
$

One more criterion, $ S_{pair}=\sum\limits_i \xi_i \cdot \xi_{i+1}$, constructed
somehow speculatively, exploits the
following idea: if the anomaly is a deviation of several neighbour
points to one side of the fitting curve (Fig. \ref{fig:pairwise}) than it
will increase the value of the statistics $S$. Thus the pairwise neighbours'
correlations test can be used as a criterion for rather general
class of anomalies.

\begin{figure}[h]
\begin{center}
\begin{minipage}[h]{0.30\linewidth}
\includegraphics[width=1\linewidth]{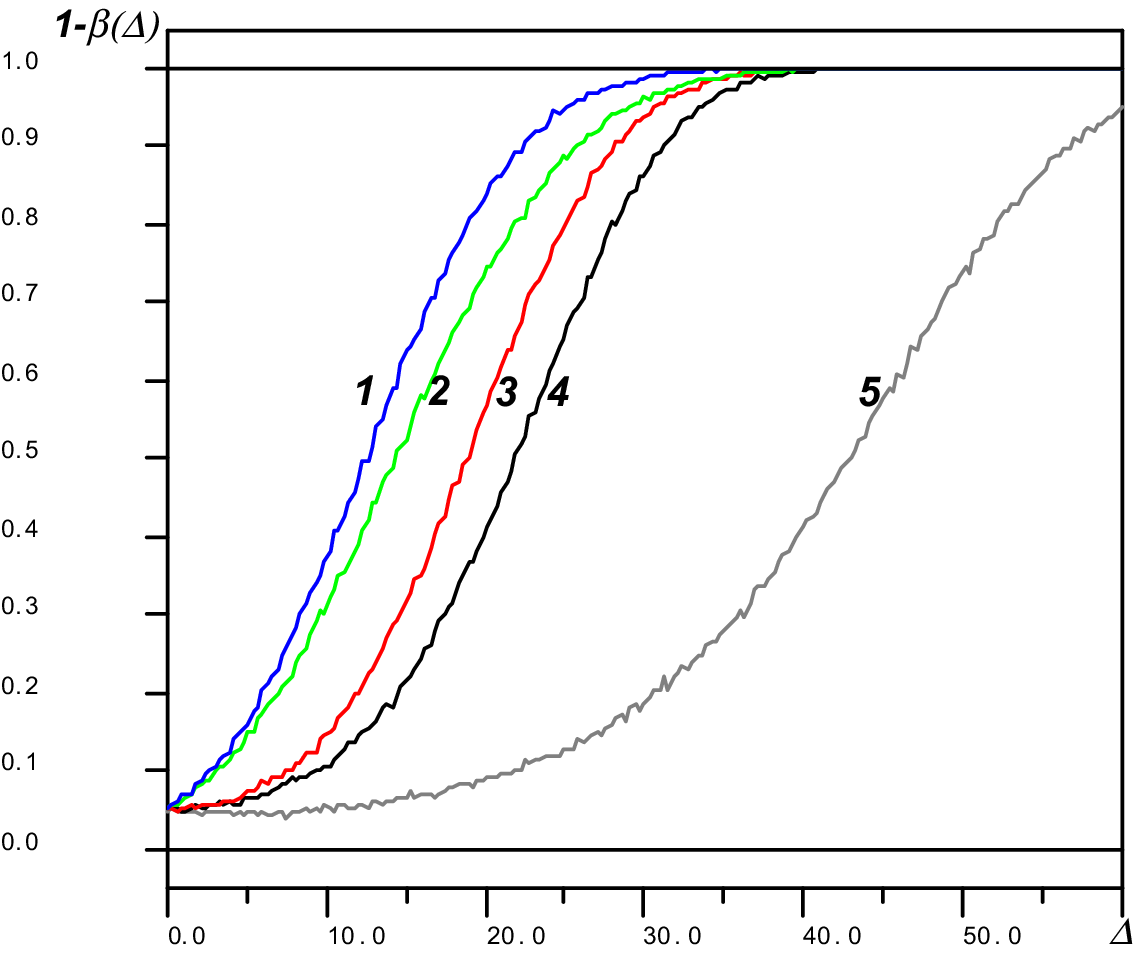}
\caption{The power functions for the special and conventional criteria} 
\label{fig:power} 
\end{minipage}
\hfill
\begin{minipage}[h]{0.62\linewidth}
\includegraphics[width=1\linewidth]{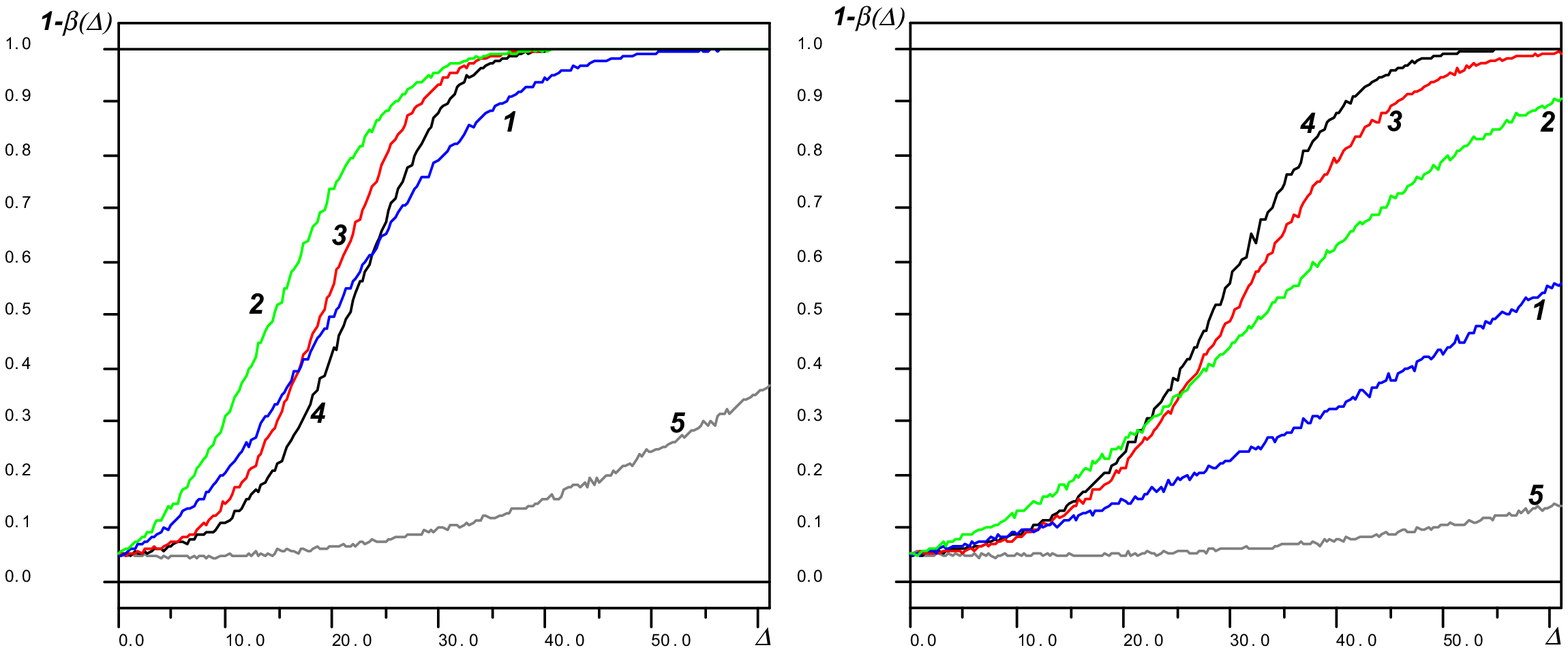}
\caption{ The power functions of the five criteria for the cases when the actual
step position is shifted from the assumed position $E_m$ by 12 eV (left) and 25
eV (right)}
\label{fig:power_shift}
\end{minipage}
\end{center}
\end{figure}

We can compare all these criteria using the standard tool of Power
Functions. The power function is simply the probability of a criterion to
reject the null hypothesis while it is in fact false.
As on can see on Fig. \ref{fig:power} the LMP (1) is the best here, the
quasi-optimal (2) is slightly less powerful, the pairwise neighbours
correlations test (3) comes third and the
conventional tests (4,5) are the least sensitive.
The situation changes if the assumed position of the step is not correct.
The left graph in Fig. \ref{fig:power_shift} shows that the LMP test (1) is
loosing its sensitivity rather rapidly while two other special tests remain
rather powerful.

\section{Heavy neutrino searches}
Similarly to the case of step-like anomaly one can derive special criteria
for the search of a heavy neutrino in $\beta$-decay spectrum
$$\frac{d\Gamma}{dE}=\sin^2\theta {\left(\frac{d\Gamma}{dE}\right)}_{m_{keV}}+
 \cos^2\theta {\left(\frac{d\Gamma}{dE}\right)}_{m_{light}}.$$ 
The spectrum $\frac{d\Gamma}{dE}(E_i,m)=S_{i,m}$ is again measured in a number
of points with various retarding potentials. The spectrum is defined by the
mixing parameter $U^2=\sin^2\theta$ and the mass of the heavy neutrino
$m_{keV}$. The mass of the light neutrino is considered to be $m_{light}=0$.
Thus the theoretical means are $\mu^{\prime}_i =
U^2S_{i,m_{H}}+(1-U^2)S_{i.m=0}$.
The first test (the LMP) is obtained via the method of quasi-optimal weights and
it is by construction the most sensitive one in case when the mass of the heavy
neutrino is well-known.
The corresponding weights are as follows:
$$
 \omega_i=\frac{\partial \ln f_i}{\partial U^2}=
 {\left(
 \frac{N}{U^2 S_{i,m_{H}}+(1-U^2)S_{i,m=0}}-1
 \right)} \cdot (S_{i,m_{H}}+S_{i,0}).
$$
Using these weights one constructs an equation
$\frac{1}{M}\sum\limits_{i=1}^{M}\omega_i=0$ for the mixing parameter estimate
$\widehat{U^2}$ -- the statistics of the LMP criterion.

To reduce the sensitivity of our test to the mass
we modify the weights (for instance, as shown in Fig. \ref{fig:q-opt2}) to
obtain the quasi-optimal criterion $ S_{q-opt}= \sum\limits_{i}^{k} w_i \cdot
\xi_i $. It is more robust and require no information about the exact mass of
the additional neutrino. The universal pairwise neighbours’ correlatiosn test
can be exploited in the case of kink searches as well.

\section{Conclusions}
We illustrated the new approach to the search for anomalies in experimental
spectra with account for the parameters with uncertainties (the position
of the step-like anomaly and the mass of the additional neutrino in our
examples). We showed that the Locally Most Powerful criterion for each anomalous
contribution can be constructed via the method of quasi-optimal moments.
Than the LMP test can be tuned to reduce the influence of the
unknown parameters of the spectra. With the help of the power functions the
constructed criteria are proved to be more efficient in searches for the
specific anomalous contributions.
The next step is to compare the sensitivity of the constructed tests
with the wavelet analysis \cite{MerArx2014:wavelet},
direct fitting and search for the kink with filters \cite{MerArx2014}.
The approach appears to be useful for the future searches of the heavy neutrino
in Troitsk \cite{AbdArx14}, \cite{BelJPG2014} and Karlsruhe \cite{MerArx2014}.

One of the authors (A.L.) would like to thank Christian Sander and Alexander
Schmidt for the invitation to participate in PANIC14, Matthias Kasemann,
Kristina Price and all the organising committee for their fantastic job in
organising this conference.

The work has been supported by RFBR grant 14-02-31055.


\begin{footnotesize}




\begin{thebibliography}{99}
\bibitem{LobNPA2003} V.M.~Lobashev, Nucl. Phys. A. {\bf 719} 153
(2003).
\bibitem{AsePRD2011} V.N.~Aseev {\it et~al.}, Phys. Rev. D. {\bf 84} 112003
(2011).
\bibitem{MerArx2014} S.~Mertens {\it et~al.}, arXiv:1409.0920 (2014).
\bibitem{MerArx2014:wavelet} S.~Mertens {\it et~al.}, arXiv:1410.7684 
(2014).
\bibitem{TkaArx06} F.V.~Tkachov, arXiv:physics/0704127 (2006).
\bibitem{LokTkaTruNIMA2012} A.V.~Lokhov, F.V.~Tkachov, P.S.~Trukhanov, Nucl.
Instrum. Meth. A{\bf 686} 162 (2012).
\bibitem{LokTkaTruNPA2013} A.V.~Lokhov, F.V.~Tkachov, P.S.~Trukhanov, Nucl.
Phys. A {\bf 897} 218
(2013).
\bibitem{AbdArx14} D.~Abdurashitov {\it et~al.}, arXiv:1403.2935 (2014).
\bibitem{BelJPG2014} A.I.~Belesev {\it et~al.}, J. Phys. G: Nucl. Part. Phys
{\bf 41} 015001 (2014).

\end{thebibliography}
%

\end{footnotesize}


\end{document}